 \documentclass[prb,twocolumn,citeautoscript,showpacs,amsmath,amssymb]{revtex4}

\usepackage[dvips]{graphicx}
\usepackage{dcolumn}
\usepackage{bm}
\usepackage{times}

\begin{document}

\title{
Electron transport through single conjugated organic molecules: \\
Basis set effects in {\it ab initio} calculations
}

\author{ San-Huang Ke,$^{1}$ Harold U. Baranger,$^{2}$ and Weitao Yang$^{1}$}

\affiliation{
     $^{\rm 1}$Department of Chemistry, Duke University, Durham, NC 27708-0354 \\
     $^{\rm 2}$Department of Physics, Duke University, Durham, NC 27708-0305
}

\date{May 22, 2007}

\begin{abstract}
We investigate electron transport through single conjugated molecules - including benzenedithiol, oligo-phenylene-ethynylenes of different lengths, and a ferrocene-containing molecule sandwiched between two gold electrodes with different contact structures - by using a single-particle Green function method combined with density functional theory calculation. We focus on the effect of the basis set in the {\it ab initio} calculation. It is shown that the position of the Fermi energy in the transport gap is sensitive to the molecule-lead charge transfer which is affected by the size of basis set. This can dramatically change, by orders of magnitude, the conductance for long molecules, though the effect is only minor for short ones. A resonance around the Fermi energy tends to pin the position of the Fermi energy and suppress this effect. The result is discussed in comparison with experimental data. 
\end{abstract}

\maketitle

\section{Introduction}

Electron transport through single conjugated organic molecules has attracted considerable experimental attention \cite{Reed97252,Reichert034137,Kubatkin03698,Rawlett023043,Zhou97611,Kushmerick03897,Xiao04267,Chen05503,He051384,Xiao059235,Getty05241401,Selzer0561} because of the convenience of the chemical assembly and their delocalized electronic states and the resulting small HOMO-LUMO gap which is very useful in molecular electronics especially for the operation under low bias voltages. Theoretically, to understand the transport properties of junctions bridged by these molecules and to be able to calculate  their conductance are important in view of both applications and fundamental physics. Regarding the comparison between theory and experiment,  however, large discrepancies have been found for some conjugated molecules between theoretical predictions \cite{Xue014292,DiVentra00979,Xue03115407,Ke05074704,Taylor03121101} and experimental conductances \cite{Reed97252,Reichert034137,Kubatkin03698,Rawlett023043,Zhou97611,Kushmerick03897}. Many efforts have been made on the theoretical side \cite{Evers235411,Burke146803,Sai186810,Toher05146402,Koentopp121403,Muralidharan155410} to understand this discrepancy and several possible reasons have been discussed. For example, the self-interaction error (SIE) and the underestimation of the transport gap in density functional theory (DFT) based {\it ab initio} calculations \cite{Toher05146402,Ke0609637}, and the neglect of the dynamical correlation effect in the non-interacting Landauer formalism \cite{Sai186810,Koentopp121403}. 


In this paper, we investigate a fundamental technical issue in all {\it ab initio} transport calculations: the effect of the basis set. We find that this effect can also cause a significant change, up to orders of magnitude, in the resulting conductance, depending on the molecular feature and the molecule-lead coupling strength. It is shown that for conjugated organic molecules with a small HOMO-LUMO gap the position of the Fermi energy ($E_F$) of the leads in the transport gap, which is determined by the molecule-lead charge transfer, is sensitive to the size of basis set. For long molecules different positions of $E_F$ can lead to significant change in the conductance, though for short molecules and strong coupling the effect may be only minor. We further show, by considering different molecule-lead contact structures, that a resonance around the Fermi energy tends to pin the position of $E_F$ and thus suppress the basis set effect, indicating that this effect will be much less important for the strong coupling limit which gives a large equilibrium conductance.

\section{Computation}

The conjugated molecules considered in our calculation include a short benzenedithiol (BDT)
molecule, oligo-phenylene-ethynylenes (OPE) of different lengths, and a
ferrocene-containing
(FC) molecule, sandwiched between two gold electrodes, as shown in Figs.~1 and 5.
We consider two models for the lead, a thin Au(001) wire as shown in Fig.~1 and a 5$\times$5
Au(111) surface as shown in Fig.~3; Both models are possible to occur in 
experiments \cite{Xiao04267,Tao053260,Li062135,Ohnishi98780,Yanson98783,Nazin0377,  
Yang02553,Melosh03112,Beckman045921,   
Datta972530,Cui01571}. We also consider two different molecule-lead contacts, through a
flat Au surface or through an apex Au atom, both of which are also possible to occur in
experiments \cite{Datta972530,Cui01571,  
Xiao04267,Tao053260,Li062135,Ohnishi98780,Yanson98783}.

\begin{figure*}[tb]
(a) \includegraphics[angle=-90,width=14.0cm]{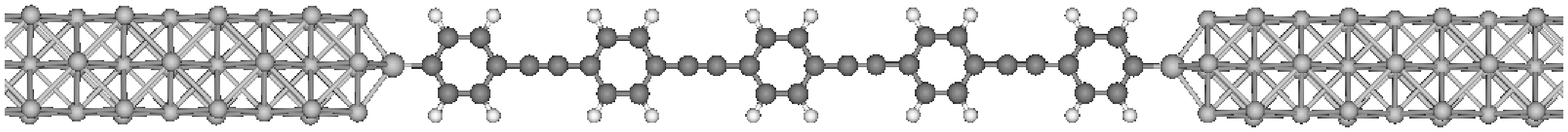} \\
(b) \includegraphics[angle=-90,width=14.0cm]{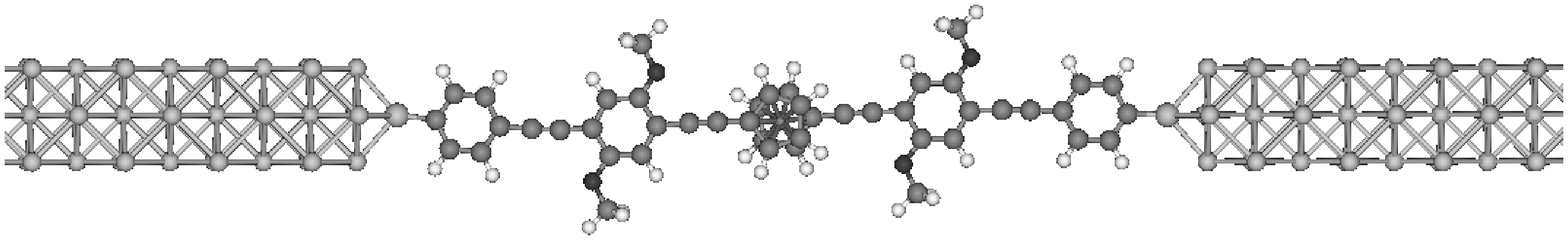}
\caption{
Structures of two molecular junctions: (a) Au(001)-OPE5-Au(001) and (b)
Au(001)-FC-Au(001). Only the extended molecule (device region) is shown, 
which contains 8 atomic layers of the lead surfaces. The lead is a
thin $\sqrt{2}\times\sqrt{2}$ Au(001) wire. 
}
\end{figure*}

We adopt a single-particle Green function method \cite{Datta95,Ke04085410} combining a Landauer formula with {\it ab initio} DFT \cite{Parr89} electronic structure calculation to investigate the electron transport through the junctions \cite{Datta95,Haug96,Ke04085410}. In practice, we divide an infinitely long Au-molecule-Au system into three parts: left lead, right lead, and device region which contains the molecule and large parts of the Au leads (see, for instance, Fig.~1) for accommodating fully the molecule-electrode interaction. The self-consistent Kohn-Sham Hamiltonian of the device region and the self-energies from the two semi-infinite Au leads (the two subsystems are treated on the same footing) are used to construct a single-particle Green function from which the transmission coefficient at any energy is calculated. The conductance, $G$, then follows from a Landauer-type relation. The detailed computational techniques have been described previously \cite{Ke04085410}. For the DFT electronic structure calculation, we make use of the optimized Troullier-Martins pseudopotentials \cite{Troullier911993} for the atomic cores. The PBE version of the generalized gradient approximation (GGA) \cite{Perdew963865} is used for the electron exchange and correlation. The wavefunctions are expanded using a numerical basis set \cite{Soler022745} of different sizes: single zeta plus polarization (SZP), double zeta plus polarization (DZP), triple zeta plus double polarization (TZDP), and triple-zeta plus triple polarization (TZTP). The atomic structure of the junctions, including the molecular structure and the first two layers of the lead surface, and the molecule-Au separation, are fully optimized by minimizing the atomic forces on the atoms to be smaller than 0.02 eV/{\AA} using the DZP basis set. Possible differences from this optimized configuration in experiments are not considered.

\section{Results and discussion}

To avoid a computational cost which is too large for the large size of basis set, we first consider the thin  
$\sqrt{2}\times\sqrt{2}$ Au(001) nanowire model for the lead, as shown in Fig.~1. Eight atomic
layers of the lead surface are included in the extended molecule to ensure 
good convergence for the molecule-lead interaction. The transmission function 
and the equilibrium conductance for the two junctions bridged by the OPE5 and FC 
molecules are given in Figs.~2 (a) and (b), respectively \cite{previous1}.
As can be seen, for both systems varying the size of the basis set changes the
position of $E_F$ of the lead in the molecular HOMO-LUMO gap.
This position is determined by the molecule-lead charge transfer; For
long-chain molecules the charge transfer will be around the contact region and
form a contact dipole. Different sizes of basis set will affect the amount of
this molecule-lead charge transfer, until it reaches a converged value. 
For usual ground-state energy a small change in the charge transfer will 
have only a minor effect. However, for molecular conductance,  
this can have a significant effect because it depends
critically on the position of $E_F$ in the transport gap, which is
very sensitive to the charge transferred - the charge in the tail of the broadened 
HOMO or LUMO state. 

\begin{figure}[tb]
\includegraphics[angle=-0,width=8.0cm]{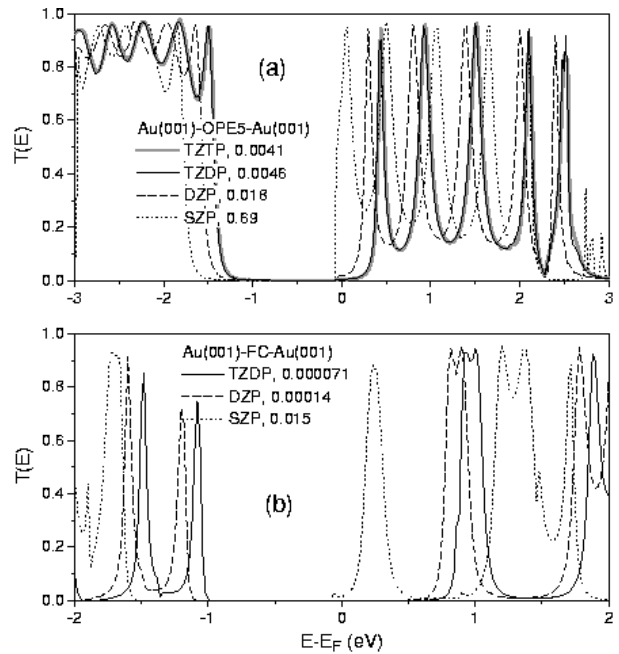}
\caption{
Transmission functions of the two molecular junctions shown in Fig.~1: (a) Au(001)-OPE5-Au(001) and (b) Au(001)-FC-Au(001). Different basis sets and the resulting equilibrium conductance (in units of $G_0$=2$e^2$/$h$) are indicated in the legend. 
}
\end{figure}

For the OPE5 junction, the SZP basis set leads to $E_F$ entering into the LUMO state and, therefore, a very large equilibrium conductance. The DZP basis set begins to put $E_F$ near the tail of the LUMO state and reduces the conductance by about a factor of 40. When we further go to the TZDP and TZTP basis sets, $E_F$ goes further away from the LUMO state and the conductance decreases further. The almost identical results for the TZDP and TZTP basis sets indicate that they have reached the converged limit. {\it Qualitatively, the DZP basis set already gives a good result though it is not good enough quantitatively.}

The result for the FC molecule [Fig.~2 (b)] shows the same trend: The SZP basis set leads to $E_F$ entering into the broadened LUMO state and a large conductance while the DZP and TZDP basis sets put $E_F$ in the middle of the gap and a much smaller conductance. From SZP to TZDP (TZTP) the relative change in conductance is more than two orders of magnitude -- an effect which has not be found for any other property.

\begin{figure}[tb]
\includegraphics[angle=-0,width=5.0cm]{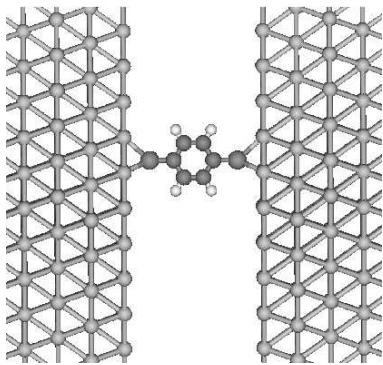}
\caption{
Structure of the Au(111)-BDT-Au(111) junction where the lead is a 5$\times$5 Au(111) surface. Only the extended molecule  (the device region) is shown, which contains 4 atomic layers of each lead surface. The structures of the other junctions bridged by OPE3, OPE5, and FC molecules are similar and are not shown here.
}
\end{figure}

\begin{figure}[tb]
\includegraphics[angle=-0,width=8.0cm]{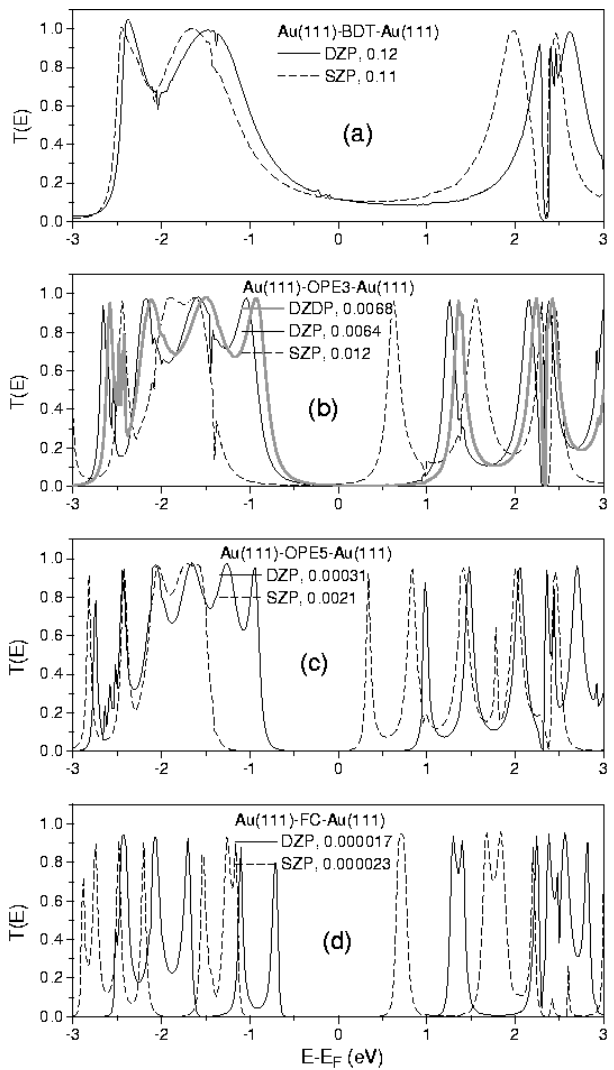}
\caption{
Transmission functions of the four junctions with the leads consisting of 5$\times$5 Au(111) surfaces: (a) Au(111)-BDT-Au(111), (b) Au(111)-OPE3-Au(111), (c) Au(111)-OPE5-Au(111), and (d) Au(111)-FC-Au(111). Basis sets used and the resulting equilibrium conductance (in units of $G_0$=2$e^2$/$h$) are indicated in the legend. 
}
\end{figure}

\begin{figure}[tb]
\includegraphics[angle=-0,width=8.3cm]{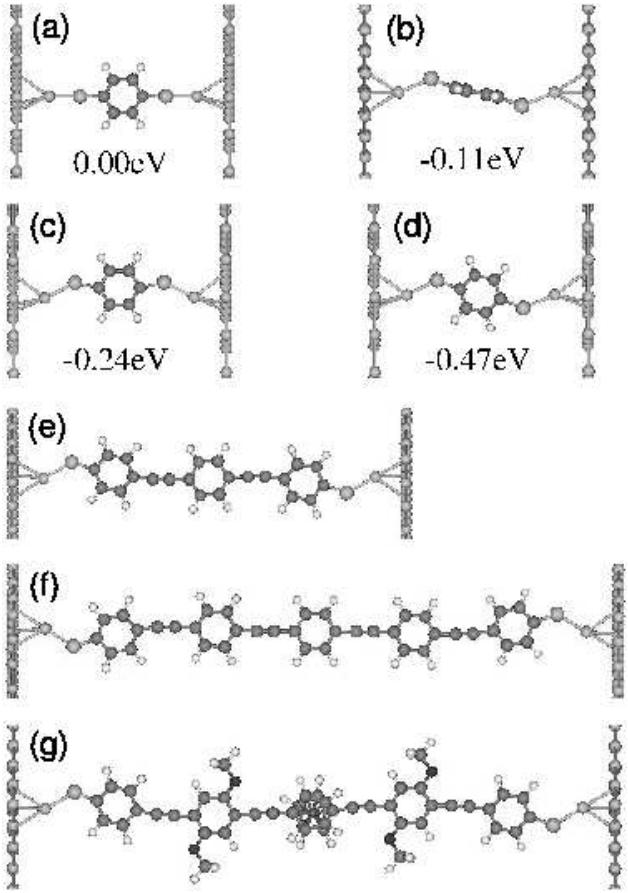}
\caption{
Structures of the junctions bridged by (a)-(d) BDT, (e) OPE3, (f) OPE5, and (g) FC molecules, where the lead is a 5$\times$5 Au(111) surface and the molecule-lead connection is through an apex Au atom. Only the molecule and the first atomic layer of the lead surface are shown. For the BDT molecular bridge, different metastable configurations are shown in (b)-(d) with the energy difference from the straight connection, (a), listed. For the OPE3, OPE5, and FC molecular bridges, only the most stable configuration is shown, which is similar to the BDT case, shown in (d).
}
\end{figure}

Next we investigate the basis set effect for conjugated molecular bridges with different lengths and also for a more realistic Au(111)-surface electrode. We consider two large 5$\times$5 Au(111) periodic surfaces as the leads bridged by the following molecules of different lengths: BDT, OPE3, OPE5, and the FC molecule. To reduce the computational cost here we consider only the SZP and DZP basis sets and include only four atomic layers of the Au(111) surface in the extended molecule, as shown in Fig.~3.   To check the convergence with the basis set we also carry out a DZDP calculation for the OPE3 system.

From the calculated transmission functions shown in Fig.~4 one can see the same effect as we discussed previously: the two different sizes of basis set affect the molecule-lead charge transfer and cause a shift of about 0.5 eV for the position of $E_F$ in the HOMO-LUMO gap. The use of the larger DZDP basis set in the test calculation for the OPE3 system leads to only a further $\sim$ 0.1eV shift [see Fig.~4 (b)] and the change in the equilibrium conductance is minor, indicating that the DZP results are almost converged. For the BDT, OPE3, and OPE5 series, the equilibrium conductance decreases along with the increasing molecule length. For the longest one, OPE5, the basis set effect causes about one order of magnitude difference in the conductance while for the shorter one, OPE3, the difference is only a factor of 2, and for the shortest one, BDT, it becomes minor. This behavior is a result of the observation that along with the increase of the molecule length the smaller SZP basis set tends to put $E_F$ closer and closer to the LUMO state while the larger DZP basis set tends to put it closer and closer to the middle of the gap. {\it As a result, the difference between the two basis sets grows as the molecule gets longer.} 

Note that one can see an approximate exponential decay of the conductance as the molecule gets longer: from BDT to OPE3, and from OPE3 to OPE5, the conductance decreases approximately by a factor of 20. This is a natural behavior of coherent tunneling and the decay ratio, $\beta$, is a property of the molecule. Here the $\beta$ value is about 0.21 {\AA}$^{-1}$ which is very close to an available experimental datum, 0.22 {\AA}$^{-1}$, for a conjugated molecule, carotenoid polyenes \cite{He051384}. The agreement between theory and experiment for the exponential decay in the case studied here is similar to that found in other cases. For instance, experimental \cite{Wold025886} and theoretical \cite{Kaun03121411} values of $\beta$ for another conjugated molecule, oligophenylene, are found to be 0.35-0.5 {\AA}$^{-1}$. Also, for alkane and peptide chains, $\beta$ is in the range 0.7-0.9 {\AA}$^{-1}$ experimentally \cite{Xiao045370,Li062135} and theoretically \cite{TomfohrSankey02,SeminarioYan05}. Naturally, the conductance of conjugated oligomers decays more slowly with chain length than that of saturated molecules; thus, the values of $\beta$ for conjugated chains are considerably smaller than those for saturated chains.

In the case of the FC molecule the situation is different, where the SZP basis set puts the $E_F$ slightly closer to the LUMO while the DZP puts it slightly closer to the HOMO, and as a result of this symmetry, the two resulting conductances   become quite close. Thus in general, the strong system dependence of the basis set effect indicates that it can cause unpredictable numerical errors which are specific to the system investigated. However, this effect can be expected to be small for short molecules and strong coupling, like for the Au-BDT-Au system,  where the strong coupling allows lead states tunneling into the molecule to meet in the middle (sometimes called metal-induced gap states), and as a result, the transmission coefficient is fairly large and flat in the gap, as is shown in Fig.~4(a).  

\begin{figure}[tb]
\includegraphics[angle=-0,width=8.0cm]{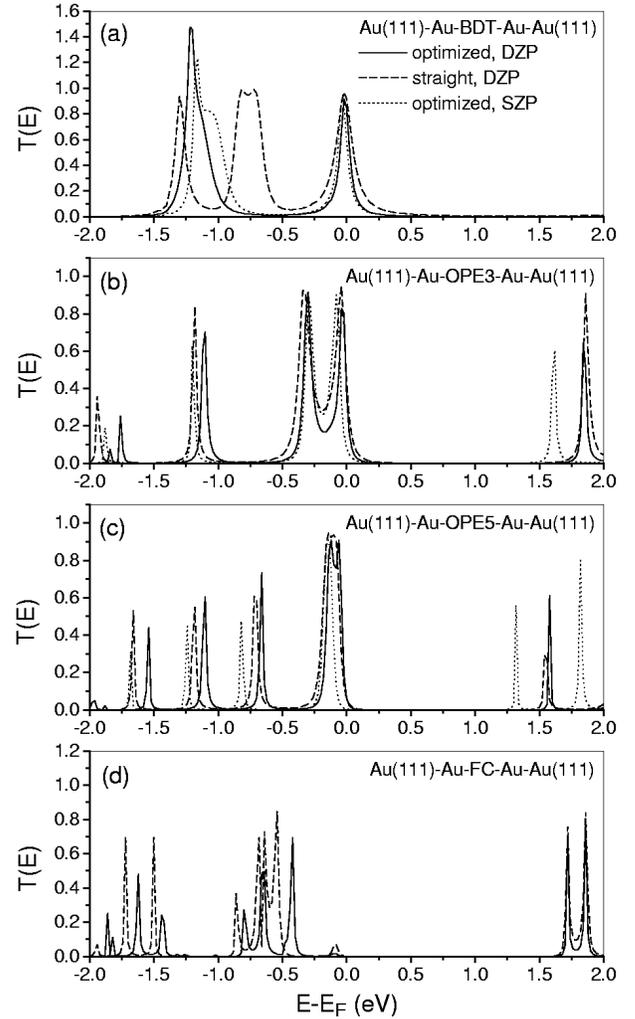}
\caption{
Transmission functions of the four junctions with the leads consisting of 5$\times$5 Au(111) surfaces, where the molecule-lead connection is through an apex Au atom: (a) Au(111)-au-BDT-au-Au(111), (b) Au(111)-au-OPE3-au-Au(111), (c) Au(111)-au-OPE5-au-Au(111), and (d) Au(111)-au-FC-au-Au(111). Results for both the straight connection and the optimized connection, as given in Fig.~5, are shown. Basis sets used are indicated in the legend. Note the resonance peak around the Fermi energy for the BDT, OPE3, and OPE5 molecule series.
}
\end{figure}

\begin{figure}[tb]
\includegraphics[angle=-0,width=8.0cm]{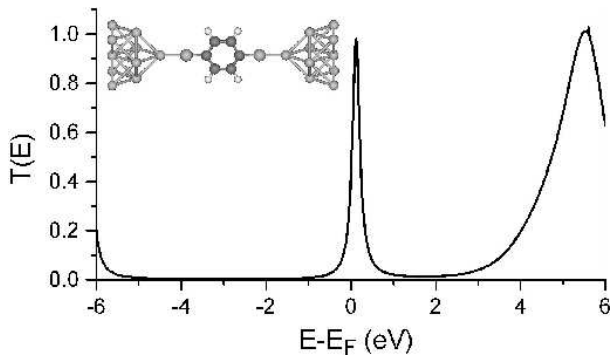}
\caption{
Transmission function of the Au(001)-Au-BDT-Au-Au(001) junction (Its structure
is shown in the inset) given by a Hartree-Fock calculation, as discussed in the
text.
}
\end{figure}

In the following we consider a possible experimental situation in which the molecule-gold connection is through an apex gold atom, as suggested by a STM experiment \cite{Xiao04267,Tao053260,Li062135}. We consider two possibilities: (1) the connection is straight with an optimized Au-S bond length, as shown in Fig.~5 (a) for a BDT molecule, which is possible to occur when the STM tip is pulled from the gold surface, as suggested in Refs.\,\onlinecite{Xiao04267}, \,\onlinecite{Tao053260}, and \,\onlinecite{Li062135}; (2) the connection and the molecule-lead separation are optimized, as shown in Figs.~5 (d)-(g). The structure optimization calculation for Au(111)-Au-BDT-Au-Au(111) junction shows three meta-stable structures, as shown in Figs.~5 (b), (c), and (d), which lower the total energy with respect to the straight connection in Fig.~5(a). The most stable structure is given in Fig.~5(d). We further find that for the longer molecules, OPE3, OPE5, and FC, the most stable optimized contact structure is similar to that in Fig.~5(d). Therefore, we only give these optimized structures while omiting those with a straight connection in Fig.~5 for the longer molecules.

The transmission functions for the four junctions are given in Fig.~6. There are several features to notice in the result. First, the presence of the apex gold atom at the contacts changes totally the transport property of the junctions. The very narrow peaks indicates that the molecule-lead coupling is significantly weakened as compared to the case without the apex gold atom. This will significantly reduce the current for large bias voltages. Second, the relaxation of the contact structure has only a small effect once the connection is through an apex gold atom. Third, a sharp resonance peak occurs around the Fermi energy except for the FC molecule where the peak near $E_F$ is very small [see Fig.~6 (d)].

We call this resonance peak the $new$ HOMO state hereafter. This resonance peak was found previously also by other calculations for straight-connected BDT systems \cite{Xue03115407,Ke05074704,Ke05114701}. Although the present calculation shows that it is a quite general behavior for conjugated molecules connected by an apex gold atom, it is unclear whether or not the occurrence of this resonance peak has anything to do with issues in DFT, like self-interaction error or the absence of the discontinuity in the exchange-correlation ($xc$) potential \cite{Ke0609637}. To check if it is an artifact due to the issues in DFT we perform a test transport calculation for a Au(001)-Au-BDT-Au-Au(001) junction (see the inset of Fig.~7) using Hartree-Fock (HF) theory which is self-interaction free and does not have the issue of $xc$ discontinuity. The calculation is carried out using a 6-31G* basis set for H, C, and S atoms and a CRENBS basis set for Au atom, and the result is given in Fig.~7. As can be seen, the resonance peak around the Fermi energy remains, indicating that it is a general behavior instead of an artifact.

Finally, the basis set effect is significantly reduced compared to the previous cases without the apex gold atom. The small effect of the basis set can be understood by considering the entering of the Fermi energy into the $new$ HOMO state. Because of it, a small change in the molecule-lead charge transfer cannot shift the position of the Fermi energy a large amount due to the total charge conserving.  This behavior indicates that the effect of the basis set will be much less important in the strong coupling limit which gives a large equilibrium conductance.

\section{Comparison to experimental results}

Finally, we would like to make some comparison between the present results and experimental data. The present result of equilibrium conductance for the Au(111)-BDT-Au(111) junction ($\sim$ 0.1$G_0$) is too large compared to experimental reports \cite{Reed97252,Xiao04267}. Similar results were also obtained by other DFT-GF calculations \cite{Xue014292,Xue03115407,DiVentra02189}. This overestimation is still an open problem and can be due to several possibilities: the difference in the atomic structure of the junction between the real experimental situation and the calculation; the self-interaction error; and the underestimation of the transport gap in the DFT electronic structure calculation \cite{Toher05146402,Ke0609637}. Because of the latter issue the conductance of the OPE3 molecule ($\sim$ 0.006$G_0$) and the OPE5 molecule ($\sim$ 0.0003$G_0$) are also expected to be overestimated.

On the other hand, in real experiments, when a molecule is long enough incoherent or inelastic effects may play a role or even take over \cite{Selzer0561}, and, as a result, the conductance will decay more slowly or even linearly with increasing molecule length. In such case the theoretical calculation for coherent tunneling may underestimate the conductance. In a certain range of molecule length these two errors may cancel with each other. 

For OPE3 molecule the present result of conductance is still too large compared to some experimental reports \cite{Xiao059235,Reichert02176804} while the very small conductance for OPE5 is in agreement with the experimental observation in Ref.\;\onlinecite{Getty05241401}.   However, our result of a very small conductance ($\sim$ 0.00002$G_0$) for the FC molecule differs qualitatively from the experimental result \cite{Getty05241401} which shows a good conduction \cite{previous2}. So far it is not clear how to understand this discrepancy. One possible reason is the structure difference between the experiment and calculation. For example, in the experiment the molecule-lead connection might be through an apex gold atom which will significantly increase the current for small bias voltages, as shown by our calculation [see Fig.~6 (d)]. Additionally, the scissor mode made by the ferrocene moiety (i.e., rotation of the two five-member rings with respect to each other) might also allow a higher conductance state to be realized, although this might not account for the big qualitative discrepancy \cite{Getty05241401}. Another possibility is the chemical absorption on the lead surface in the experiment \cite{Getty05241401} which was partially done in solution first. The chemical absorption may change the work function of the surface of the electrode or form a dipole layer on the surface, both of which can change the lineup between the Fermi energy and the HOMO or LUMO state and therefore cause a large increase in the conductance. 

\section{Summary}

We have investigated electron transport through different single conjugated molecules of different length by using single-particle Green function method combined with density functional theory calculation. It is shown that the convergence with respect to the size of basis set is important for {\it ab initio} transport calculations since the position of the Fermi energy in the transport gap is sensitive to the molecule-lead charge transfer which is affected by different sizes of basis set. It is shown that this basis set effect can be dramatic, up to orders of magnitude, while it is only minor for short molecules and strong coupling.

Our calculation also shows that a resonance around the Fermi energy tends to pin the position of the Fermi energy and suppress this effect, indicating that it will be much less important for the strong coupling limit which gives a large equilibrium conductance.
Comparing to experiments, our result for the OPE molecules is in agreement with some experimental reports. The result for the FC molecule is not in qualitative agreement and the reason needs further experimental and theoretical work to clarify.

This work was supported in part by the NSF (DMR-0506953).



\begin{thebibliography}{51}
\bibitem[{pre({\natexlab{a}})}]{previous1}
\bibinfo{note}{In a previous work (Ref.\,\onlinecite{Getty05241401}) we
  calculated the OPE5 and FC systems with thicker Au(001) leads adopting the
  SZP basis set. The results are qualitatively consistent with the present
  calculation using the thinner Au(001) lead and adopting the SZP basis set.}
\bibitem[{pre({\natexlab{b}})}]{previous2}
\bibinfo{note}{In a previous calculation (Ref.\,\onlinecite{Getty05241401}) we
  obtained a large equilibrium conductance for both the OPE5 and FC systems due
  to the use of a too small basis set (SZP).}
\end{thebibliography}

\end{document}